\documentclass[12pt]{article}
\usepackage{amssymb}
\usepackage[dvips]{epsfig}

\newcommand{\be}{\begin{equation}}
\newcommand{\ee}{\end{equation}}
\def\bea{\begin{eqnarray}}
\def\eea{\end{eqnarray}}

\newcommand{\bn}{\begin{eqnarray}}
\newcommand{\en}{\end{eqnarray}}

\newcommand{\p}{\partial}

\newcommand{\nn}{\nonumber}
\newcommand{\tih}{\tilde{h}}
\newcommand{\no}{\noindent}
\newcommand{\tf}{\tilde{f}}

\newcommand{\cL}{{\cal{L}}_}
\newcommand{\sdf}{S_{SD}^{(4)}}

\newcommand{\s}{\,\,\,\,}
\def\bea{\begin{eqnarray}}
\def\eea{\end{eqnarray}}

\newcommand{\beq}{\begin{eqnarray}}
\newcommand{\eeq}{\end{eqnarray}}
\begin{document}

\title{\textbf{A new spin-2 self-dual model in
$D=2+1$ }}
\author{D. Dalmazi  and Elias L. Mendon\c ca \\
\textit{{UNESP - Campus de Guaratinguet\'a - DFQ} }\\
\textit{{Av. Dr. Ariberto Pereira da Cunha, 333} }\\
\textit{{CEP 12516-410 - Guaratinguet\'a - SP - Brazil.} }\\
\textsf{E-mail: dalmazi@feg.unesp.br , elias.fis@gmail.com }}
\date{\today}
\maketitle

\begin{abstract}

There are three self-dual models of massive particles of helicity $+2$ (or $-2$) in $D=2+1$. Each model is of
first, second, and third-order in derivatives. Here we derive a new self-dual model of fourth-order, $\cL
{SD}^{(4)}$, which follows from the third-order model (linearized topologically massive gravity) via Noether
embedment of the linearized Weyl symmetry. In fact, each self-dual model can be obtained from the previous one $
\cL {SD}^{(i)} \to \cL {SD}^{(i+1)} \, , \, i=1,2,3$ by the Noether embedment of an appropriate gauge symmetry,
culminating in $\cL {SD}^{(4)}$. The new model may be identified with the linearized version of $\cL {HDTMG} =
\epsilon^{\mu\nu\rho}\Gamma_{\mu\gamma}^{\epsilon}\left\lbrack \p_{\nu}\Gamma_{\epsilon\rho}^{\gamma} +
(2/3)\Gamma_{\nu\delta}^{\gamma}\Gamma_{\rho\epsilon}^{\delta} \right\rbrack /8 m + \sqrt{-g}\left\lbrack
R_{\mu\nu} R^{\nu\mu} - 3\, R^2/8 \right\rbrack /2 m^2 $. We also construct a master action relating the
third-order self-dual model to $\cL {SD}^{(4)}$ by means of a mixing term with no particle content which assures
spectrum equivalence of $\cL {SD}^{(4)}$ to other lower-order self-dual models despite its pure higher
derivative nature and the absence of the Einstein-Hilbert action. The relevant degrees of freedom of $\cL
{SD}^{(4)}$ are encoded in a rank-two tensor which is symmetric, traceless and transverse due to trivial
(non-dynamic) identities, contrary to other spin-2 self-dual models. We also show that the Noether embedment of
the  Fierz-Pauli theory leads to the new massive gravity of Bergshoeff, Hohm and Townsend.

\end{abstract}

\newpage

\section{Introduction}

It is known that in $D=2+1$, massive particles of helicity $+1$ (or $-1$) can be described either by a second
order gauge theory \cite{djt}, Maxwell-Chern-Simons ($\cL {MCS}(A_{\mu})$), or by a first-order nongauge theory
\cite{tpn}, self-dual model
 ($\cL {SD}(f)$). The physical equivalence of both theories can be
 established via a master action \cite{dj} depending on both fields $A_{\mu}$ and $f_{\mu}$ which
 is obtained from the self-dual model $\cL {SD}(f)$ by adding a mixing term
 between the fields $A_{\mu}$ and $f_{\mu}$. Since the mixing term
 is a pure first-order Chern-Simons term $CS_1$ with no particle
 content, the physical equivalence between $\cL {MCS}(A_{\mu})$
 and $\cL {SD}(f)$ follows trivially. In particular, this
 explains why the propagator of the MCS theory contains an innocuous
 (vanishing residue \cite{baeta}) massless pole besides the
 physical massive pole present in the self-dual model of
 \cite{tpn}. Namely, the non-propagating massless pole is inherited from the pure
 Chern-Simons term. Alternatively, one can derive the MCS theory
 out of $\cL {SD}(f)$ via a two steps Noether embedment
 ,see \cite{anacleto}, of the gauge symmetry $\delta_{\Lambda}f_{\mu}=\p_{\mu}\Lambda
 $ of the Chern-Simons term present in
  $\cL {SD}(f)$. Since a couple of parity singlets of opposite helicities $+1$ and $-1$ can be
  combined into one parity doublet (Proca theory), one might try
  to apply the Noether gauge embedment (NGE) procedure directly to
  the Proca model. Indeed, in the begin of the next section, as an
  introduction to the rest of the work, we show that in this case
  we obtain a (``wrong'' sign) Maxwell-Podolsky theory which
  contains a massive physical particle plus a massless ghost in
  the spectrum. Thus, the NGE procedure, in this case, fails to
  produce a physical gauge theory. The appearance of ghosts via NGE has been noticed before
  in  \cite{baeta}. The analogous of the Proca model for spin-2
  particles is the Fierz-Pauli theory. In the next section we show
  that in this case the NGE procedure leads, in $D=2+1$, precisely
  to the new massive gravity theory of Bergshoeff, Hohm and
  Townsend (BHT model henceforth) \cite{bht}. Such theory shares
  the same spectrum of the Fierz-Pauli theory. We explain the difference
  between the spin-1 and spin-2 cases based on the different
  particle contents of the Einstein-Hilbert and Maxwell actions.

  In the third and main section we apply the NGE procedure to
  parity singlets of helicity $+2$. We show that the three known
  self-dual models of first- ($S_{SD}^{(1)}$), second-
  ($S_{SD}^{(2)}$) and third-order ($S_{SD}^{(3)}$), see
  respectively \cite{aragone,desermc,djt},  can be related
  ($S_{SD}^{(1)} \to S_{SD}^{(2)} \to S_{SD}^{(3)}$) via Noether
  embedment of appropriate gauge symmetries, which is in agreement
  with the triple master action of \cite{prd2009}. In subsection 3.3, by embedding
  a linearized Weyl symmetry present in part of $S_{SD}^{(3)}$ (linearized
  topologically massive gravity (LTMG)) we obtain a previously
  unknown fourth-order self-dual model ($S_{SD}^{(4)}$) dual
   to the other self-dual models which completes the
  sequence of embedment with $S_{SD}^{(3)} \to S_{SD}^{(4)}$. In
  section 4 we draw our conclusions.

\section{Gauge embedment of parity doublets}

\subsection{The spin-1 case}

It is known that massive particles of spin-1 are described in a
covariant way by the Proca model:

\be {\cal{L}}_P= - \frac{1}{4}F^{\mu\nu}F_{\mu\nu} -
\frac{m^2}{2}A^{\mu}A_{\mu}.\label{proca} \ee

\no Throughout this work we use, $\mu,\nu =0,1,2$ and the signature is $\eta_{\mu\nu}=(-,+,+)$. From the
equations of motion of (\ref{proca}) one derives the transverse condition $\p_{\mu}A^{\mu}=0$  and the
Klein-Gordon equation $(\Box - m^2)A^{\mu}=0$. The Lagrangian (\ref{proca}) contains a parity doublet of
helicities $+1$ and $-1$ in $D=2+1$, for a simple derivation see \cite{solda2}. The Maxwell term is invariant
under the gauge transformation $\delta_{\Lambda} A_{\mu} = \p_{\mu} \Lambda$ which is broken by the mass term.
One might wonder whether there would exist a gauge invariant description of  spin-1 massive particles. Let us
show how does the Noether gauge embedment procedure \cite{anacleto} work in this case. The gauge variation of
the Proca action, $S_P=\int d^3x\,\cL P$, can be written as:

\be \delta_{\Lambda} S_{P} = \int d^3x\, K^{\nu}\p_{\nu}\Lambda
\label{dsp}\ee

\no The Euler vector is given by $K^{\nu}=\delta S_P/\delta A_{\nu}=(\Box\theta^{\nu\mu}-m^2g^{\nu\mu})A_{\mu}$
with $\Box\theta^{\nu\mu}=\Box\eta^{\mu\nu}-\p^{\mu}\p^{\nu}$. As a first step in the Noether procedure one
introduces a compensating auxiliary vector field whose gauge transformation is given by $\delta_{\Lambda}
a_{\nu} = -\p_{\nu}\Lambda $ such that:

\be \delta_{\Lambda} S_1\equiv \delta_{\Lambda} \left( S_P + \int d^3x\, K^{\nu} a_{\nu} \right)= \int d^3x\,
\delta_{\Lambda} K^{\nu} a_{\nu} = \int d^3x\,\left( - m^2 \p^{\nu}\Lambda \, a_{\nu} \right) = \delta_{\Lambda}
\, \int d^3x \left( \frac{m^2}2 a^{\nu} a_{\nu} \right) \label{ds1}\quad . \ee

\no Therefore,

\be \delta_{\Lambda} S_2 \equiv  \delta_{\Lambda} \int d^3x\,
\left( \cL P + K^{\nu} a_{\nu} - \frac{m^2}2 a^{\nu} a_{\nu}
\right) = 0 \label{ds2} \quad . \ee

\no Eliminating the auxiliary field $a_{\mu}$ by means of its equations of motion from the second iterated
action $S_2$ defined above we end up with the higher-order gauge invariant action:

\be S_{{\rm inv.}} = \int d^3x\, \left( \cL P
+\frac{K^{\nu}K_{\nu}}{2m^2} \right) = \frac 14 \int d^3x\,
F^{\mu\nu}\left( 1 - \frac{\Box}{m^2} \right)F_{\mu\nu}
\label{podolsky} \ee

\no The addition of a quadratic term in the Euler vector to the Proca theory guarantees that an arbitrary
variation $\delta S_{{\rm inv.}}= \int d^3 x \, K^{\nu}\left(\delta A_{\nu} + \delta K_{\nu}/m^2 \right) $ will
vanish at $K_{\nu}=0$. So the solutions of the equations of motion of the original Proca theory will be also
solutions of the equations of motion of the new action $ S_{{\rm inv.}}$. Thus, the Proca theory is embedded in
the gauge theory (\ref{podolsky}) which is the three dimensional analogue of the Podolsky \cite{podolsky} theory
but with an opposite overall sign. The equations of motion of $S_{{\rm inv.}}$, i.e., $\left(\Box - m^2
\right)\p^{\mu}F_{\mu\nu} = 0$, in the gauge $\p^{\mu}A_{\mu}=0$, lead to $\Box \left(\Box - m^2\right)A_{\mu}
=0$. So, besides the expected massive particle we have also a massless mode. The overall sign in
(\ref{podolsky}) is such that the massive particle is physical and the  massless one is a ghost in agreement
with \cite{baeta}. This is contrary to the Podolsky theory which is known to contain a massless photon and a
massive ghost, see comment in \cite{accioly}. In summary, we have not succeeded in deriving a physical gauge
theory for spin 1 particles by a direct embedment of (\ref{proca})\footnote{If we linearize the Proca theory by
introducing an auxiliary vector field we do derive a physically consistent gauge model dual to the Proca theory
which is the $D=2+1$ version of the Kalb-Ramond theory.}. This can be better understood from the master action
point of view. In the master action approach for massive theories \cite{dj,jhep1,jhep2}, we add to the original
non-gauge theory, the Proca model, a mixing term between the dual fields  with the desired gauge symmetry such
that the highest derivative term of the non-gauge theory is canceled. In the present case we are led to the
Master action:

\be S_M(A,\tilde{A}) = \int d^3x
\left\lbrack-\frac{1}{4}F_{\mu\nu}(A)F^{\mu\nu}(A)-\frac{m^2}{2}
A_{\mu}A^{\mu}+\frac{1}{4}F_{\mu\nu}(A-\tilde{A})F^{\mu\nu}(A-\tilde{A})\right\rbrack
\label{sm1}\ee

\no The action (\ref{sm1}) is invariant under $\delta_{\Lambda} \tilde{A}_{\mu} = \p_{\mu} \Lambda $. Due to the
positive sign in front of the Maxwell-type mixing term the path integral over the non-gauge field $A_{\mu}$
leads to a local theory which is exactly the gauge theory (\ref{podolsky}) with $A_{\mu}$ replaced by the dual
gauge field $\tilde{A}_{\mu}$. On the other hand if we make the shift $\tilde{A}_{\mu} \to \tilde{A}_{\mu} +
A_{\mu}$ in (\ref{sm1}) before any integration we obtain a Proca theory plus a decoupled Maxwell term with
``wrong'' overall sign which is the origin of the massless ghost in agreement with \cite{baeta}.

\subsection{The spin-2 case}

 In the last subsection, we have obtained a higher order Maxwell-Podolski-type model with a ghost.
 Here we will see that the same procedure applied to
the Fierz-Pauli theory (spin-2 analogue of Maxwell-Proca model) leads us to a higher order theory without ghosts
in the spectrum, the differences will be explained in the master action context. The spin-2 higher order model,
is the linearized version of the new massive gravity suggested in \cite{bht}.

We start with the Fierz-Pauli \cite{fierz} theory, which describes in $D=2+1$ a parity doublet of massive
particles of helicities $+2$ and $-2$, see \cite{solda2} again for a simple proof. Introducing a source term we
can write this theory as follows:

\bea S_{FP}[j]&=&\int
d^3x\,\,\left\lbrack\frac{1}{2}T_{\mu\nu}(h)T^{\nu\mu}(h)-\frac{1}{4}T^2(h)-\frac{m^2}{2}(h_{\mu\nu}h^{\nu\mu}-h^2)+j_{\mu\nu}h^{\mu\nu}\right\rbrack\label{fpa}
\\ &=& \int
d^3x\,\,\left\lbrack\frac 12 \left(\sqrt{-g}
R\right)_{hh}-\frac{m^2}{2}(h_{\mu\nu}h^{\nu\mu}-h^2)+j_{\mu\nu}h^{\mu\nu}\right\rbrack
\label{fpb} \eea

\no where $\left(\sqrt{-g} R\right)_{hh}$ stands for the quadratic  truncation of the Einstein-Hilbert action in
the Dreibein fluctuations about a flat background ($ e_{\alpha\beta} = \eta_{\alpha\beta} + h_{\alpha\beta}$)
and

\be T_{\mu\nu}(h) \equiv
\epsilon_{\mu\alpha\beta}\p^{\alpha}h^{\beta}_{\,\,\nu} = -
E_{\mu\beta} h^{\beta}_{\,\nu} \quad , \label{t} \ee

\be E_{\mu\beta} \equiv \epsilon_{\mu\beta\delta}\p^{\delta}
\label{e}\ee

\no It is important to mention that throughout this work we use
second rank tensor fields, like $h_{\alpha\beta}$ in (\ref{fpa}),
with no symmetry in their indices. Symmetric and antisymmetric
combinations will be denoted respectively by:
$h_{(\alpha\beta)}\equiv \left( h_{\alpha\beta} + h_{\beta\alpha}
\right)/2 $ and $h_{\lbrack \alpha\beta\rbrack}\equiv \left(
h_{\alpha\beta} - h_{\beta\alpha} \right)/2 $.

\nn The Fierz-Pauli action leads to the following equations of motion (at vanishing sources)

\be E^{\mu\alpha}E^{\nu\beta}h_{(\alpha\beta)} = m^2 \, \left( h^{\nu\mu} - \eta^{\nu\mu} \, h \right) \,
\label{eqfp} \ee

\no from which one can derive all the required constraints to describe a spin-2 particle in $D=2+1$:

\bea
h = h_{\mu}^{\,\mu} &=& 0 \label{trace}\\
h_{\lbrack \alpha\beta\rbrack} &=& 0 \label{anti}\\
\p^{\alpha}h_{\alpha\beta} &=& 0 \,= \,\p^{\beta}h_{\alpha\beta} \label{transverse}\eea

\no as well as  the Klein-Gordon equation $\left(\Box - m^2 \right)h_{\alpha\beta}=0$.

Regarding the NGE procedure it is important to note that the Einstein-Hilbert action $\left(\sqrt{-g}
R\right)_{hh}$ is invariant under the local gauge symmetries:

\be \delta_G
h_{\mu\nu}=\p_{\mu}\xi_{\nu}+\epsilon_{\mu\nu\alpha}\Lambda^{\alpha}\label{sym0}\ee

\no which are broken by the Fierz-Pauli mass term. In order to
embed this symmetries in a new model, we  calculate the Euler
tensor from $S_{FP}[j]$:

\be M^{\mu\nu}=\frac{\delta S_{FP}\lbrack j \rbrack }{\delta
h_{\mu\nu}}=E^{\beta\nu}T^{\nu}_{\s\beta}-\frac{1}{2}E^{\nu\mu}T-m^2(h^{\nu\mu}-\eta^{\mu\nu}h)+j^{\mu\nu}\label{mmini}\ee

\no Then, we propose the following first iterated action $S_1$
 by using an auxiliary tensor field $a_{\mu\nu}$ such
that

\be \delta_G a_{\mu\nu}=-\delta_G h_{\mu\nu} \label{da} \ee

\be S_1 = \int d^3 x \, \left( \cL {FP} + a_{\mu\nu}M^{\mu\nu}+
j_{\mu\nu}h^{\mu\nu} \right) \ee

\no Following the same steps of the previous subsection, it is easy to prove that the action below, is invariant
under the gauge transformations (\ref{sym0}), (\ref{da}):

\be S_2 = \int d^3 x \, \left( \cL {FP} +
j_{\mu\nu}h^{\mu\nu}+a_{\mu\nu}M^{\mu\nu}-\frac{m^2}{2}(a_{\mu\nu}a^{\nu\mu}-a^2)\right)
\label{bhta}\ee

\no Getting rid of the auxiliary fields by means of their
algebraic equations of motion we obtain

 \bea \cL {{\rm inv.}} &=& \cL
{FP}+j_{\mu\nu}h^{\mu\nu}+
\frac{1}{2m^2}M_{\mu\nu}M^{\nu\mu}-\frac{1}{4m^2}M^2\label{bhtb}\\
&=& -\frac{1}{2}T_{\mu\nu}T^{\nu\mu}+\frac{1}{4}T^2
+\frac{1}{4m^2}h_{(\rho\sigma)}\Box^2(2\theta^{\rho\nu}\theta^{\sigma\mu}-
\theta^{\rho\sigma}\theta^{\mu\nu})h_{(\mu\nu)}+j_{\mu\nu}H^{\mu\nu}(h) \label{bhtc} \eea

\no where we have neglected quadratic terms in the sources which are not important for our purposes and

\be H^{\mu\nu}(h)=\frac{1}{m^2}\left\lbrack
-E^{\beta\mu}E^{\nu\alpha}+\frac{1}{2}E^{\nu\mu}E^{\beta\alpha}+\frac{\eta^{\mu\nu}}{2}E^{\beta\gamma}E_{\gamma}^{\s\alpha}\right\rbrack
h_{\alpha\beta} \label{fmn}\ee

\no The action (\ref{bhtc}), at vanishing sources, corresponds exactly to the quadratic truncation of the new
massive gravity recently proposed \cite{bht} up to an overall $1/2$ factor, i.e.,

\be \cL {{\rm inv.}} = \cL {BHT} (j) = \left\lbrack -\frac{\sqrt{-g}}{2} R+
\frac{1}{2m^2}\left(R_{\mu\nu}R^{\mu\nu}-\frac{3}{8}R^{2}\right)\right\rbrack_{hh}+j_{\mu\nu}H^{\mu\nu}(h)
\label{bhtd}\ee

\no The action (\ref{bhtd}) is invariant under the local symmetries (\ref{sym0}) as required. From the  linear
terms in the sources of (\ref{fpa}) and (\ref{bhtd}) we have the dual map

\be h^{\mu\nu} \leftrightarrow H^{\mu\nu}(h)\label{mapbht}\ee

\no The quadratic terms in the Euler tensors in (\ref{bhtb}) assure once again that the equations of motion of
the Fierz-Pauli theory (at vanishing sources) are embedded in the BHT equations of motion which are, at
$j_{\mu\nu}=0$,

\be \Box^2(2\theta^{\rho\nu}\theta^{\sigma\mu}-\theta^{\rho\sigma}\theta^{\mu\nu})h_{(\mu\nu)} = 2 m^2 \left(
E^{\rho\mu}E^{\sigma\nu} \right)h_{(\mu\nu)} \label{eqbht} \quad . \ee

\no In fact we can rewrite (\ref{eqbht}) in the form (\ref{eqfp}) by noting that from (\ref{fmn}) we have
$m^2(H^{\mu\nu}(h)-\eta^{\mu\nu}H(h))=E^{\rho\mu}E^{\sigma\nu}h_{(\rho\sigma)}$ and applying the operator
$E^{\rho\mu}E^{\sigma\nu}$ on (\ref{fmn}) we can verify $ E^{\rho\mu}E^{\sigma\nu}H_{\mu\nu}(h)=\Box^2
\left(2\theta^{\beta\rho}\theta^{\alpha\sigma}-\theta^{\rho\sigma}
\theta^{\beta\alpha}\right)h_{(\alpha\beta)}/2m^2 $. Then, (\ref{eqbht}) implies $
E^{\mu\alpha}E^{\nu\beta}H_{\alpha\beta} = m^2 \, \left\lbrack H^{(\nu\mu)}(h) - \eta^{\nu\mu}H(h)\right\rbrack
\, $ which, compare with (\ref{eqfp}), confirms the dual map (\ref{mapbht}) at classical level.

From the  remarks of the previous subsection one might wonder whether the higher derivative BHT theory contains
ghosts introduced by the NGE procedure. As shown in \cite{oda} this is not the case. A simple demonstration is
based on the following master action \cite{bht} which parallels (\ref{sm1}):

\be S_M\left\lbrack h, \tih \right\rbrack = \int d^3 x\,
\left\lbrack \frac 12 \left( \sqrt{-g} R\right)_{hh}
-\frac{m^2}{2}(h_{\mu\nu}h^{\nu\mu}-h^2) - \frac 12 \left(
\sqrt{-g} R\right)_{h-\tih,h-\tih}\right\rbrack \label{sm2} \ee

\no On one hand, the additional mixing term of the Einstein-Hilbert type cancels out the first term of
(\ref{sm2}) such that the integration over $h_{\mu\nu}$ becomes Gaussian which gives rise exactly to the BHT
theory (\ref{bhtd}) with $h_{\mu\nu}$ substituted by the dual field $\tih_{\mu\nu}$. On the other hand, if
instead of integrating over $h_{\mu\nu}$ we shift $\tih_{\mu\nu} \to \tih_{\mu\nu} + h_{\mu\nu}$ the ``wrong''
sign Einstein-Hilbert term decouples from the Fierz-Pauli theory. Since, contrary to the Maxwell term in the
spin-1 case, the Einstein-Hilbert term has no particle content, the spectrum of the BHT model must be the same
of the Fierz-Pauli theory which explains the success of the NGE procedure in the spin-2 case.

\section{Gauge embedment of parity singlets of spin-2}

\subsection{Embedding $S_{SD}^{(1)}$ in $S_{SD}^{(2)}$ }

In this section our starting point is the first order self-dual model of \cite{aragone} which describes a
massive particle of helicity $+2$ in $D=2+1$. Introducing a source term for future purposes we have:

 \be S_{SD}^{(1)}[j]= \int d^3x\,\,
\left\lbrack \frac{m}{2}\epsilon^{\mu\nu\lambda} f_{\mu}^{\s
\alpha}\p_{\nu}f_{\lambda\alpha}-
\frac{m^2}{2}(f_{\mu\nu}f^{\nu\mu}-f^2)+f_{\mu\nu}j^{\mu\nu}\right\rbrack
\label{lsd1} \ee

\no The equations of motion of (\ref{lsd1}) in the absence of
sources,

\be E_{\mu}^{\,\lambda}f_{\lambda\alpha} =
m\left(\eta_{\mu\alpha}\, f- f_{\alpha\mu}
 \right) \qquad , \label{eq1} \ee

\no also lead to the constraints (\ref{trace}),(\ref{anti}) and (\ref{transverse}) and the Klein-Gordon equation
$\left(\Box - m^2\right) f_{\alpha\beta}= 0$. From (\ref{eq1}) we have the helicity equation $\left(
J^{\mu}P_{\mu}/\sqrt{-P^2} + 2 \right)^{\alpha\beta\gamma\delta} f_{\gamma\delta}=0$, with
$2^{\alpha\beta\gamma\delta} = 2 \left( \delta^{\alpha\gamma} \delta^{\beta\delta} + \delta^{\alpha\delta}
\delta^{\beta\gamma}\right)$, $P_{\mu}=-i \p_{\mu}$ and $\left(J^ {\mu}\right)^{\alpha\beta\gamma\delta} = i\,
\left(\eta^{\alpha\gamma}\epsilon^{\beta\mu\delta} + \eta^{\beta\gamma}\epsilon^{\alpha\mu\delta} +
\eta^{\alpha\delta}\epsilon^{\beta\mu\gamma} + \eta^{\beta\delta}\epsilon^{\alpha\mu\gamma} \right)/2$ , see
\cite{gaitan}, which assures that we are dealing with a parity singlet of helicity $+2$.

 The local symmetry \be \delta_{\xi} f_{\mu\nu}=\p_{\mu}\xi_{\nu}\label{sim1}\ee

\no of the first term in (\ref{lsd1}) is broken by the Fierz-Pauli mass term and the source term. However,
through the Noether gauge embedment procedure, we can recover it. Repeating the procedure of last section, we
begin by computing the Euler tensor: \be M^{\beta\gamma}=\frac{\delta S_{SD}^{(1)}\lbrack j \rbrack }{\delta
f_{\beta\gamma}}=-mE^{\beta\lambda}f_{\lambda}^{\s\gamma}-m^2(f^{\gamma\beta}-
\eta^{\beta\gamma}f)+j^{\beta\gamma}\label{euler1}\ee

\no With the help of an auxiliary field which satisfies
$\delta_{\xi} a_{\beta\gamma}= - \p_{\beta} \xi_{\gamma} $ we can
define a first-iterated action

\be \delta_{\xi} S^{(1)} = \delta_{\xi} \int d^3 x\, \left(\cL {SD}^{(1)} +
a_{\beta\gamma}M^{\beta\gamma}\right)= \int d^3 x\, a_{\beta\gamma}\delta_{\xi} M^{\beta\gamma} = \delta_{\xi}
\int d^3 x\,\frac{m^2}{2}(a_{\beta\gamma}a^{\gamma\beta}-a^2),\ee

\no where we have used $\delta_{\xi} a_{\mu\nu} = -\delta_{\xi} f_{\mu\nu}$. Thus, we can obtain the gauge
invariant model:

\be \cL 2 = \cL {SD}^{(1)}-a_{\beta\gamma}M^{\beta\gamma} -\frac{m^2}{2}(a_{\beta\gamma}a^{\gamma\beta}-a^2)\ee

\no Solving the equations of motion for $ a_{\beta \gamma} $ and replacing the solutions we find, after dropping
quadratic terms in the sources,

\bea \cL {SD}^{(2)}&=&\cL {SD}^{(1)}+
f_{\mu\nu}j^{\mu\nu}+\frac{1}{2m^2}M_{\nu\mu}M^{\mu\nu}-\frac{1}{4m^2}M^2\label{sd2a}\\
&=& \frac{1}{2}T_{\mu\nu}(f) T^{\nu\mu}(f) -\frac{1}{4}T^{2}(f)-\frac m2 f_{\mu\nu} T^{\mu\nu}(f) +
j_{\mu\nu}F^{\mu\nu}(f) ,\label{sd2b}\\
&=& \frac 12 \left(\sqrt{-g} R\right)_{ff} -
\frac{m}{2}\epsilon^{\mu\alpha\beta}f_{\mu\nu}\p_{\alpha}f_{\beta}^{\s\nu}+j_{\mu\nu}F^{\mu\nu}(f) ,
\label{sd2c} \eea

\no where  $ e_{\alpha\beta} = \eta_{\alpha\beta} + f_{\alpha\beta}$ and

\be F^{\mu\nu}(f)= \frac{1}{m}\left\lbrack T^{\nu\mu}(f)-\frac{\eta^{\mu\nu}}{2}T(f)\right\rbrack \quad ,
\label{f} \ee

\no Note that the last two terms in (\ref{sd2a}) are quadratic in the Euler tensor which guarantees again the
embedment of the equations of motion of $\cL {SD}^{(1)}$ in the second-order model $\cL {SD}^{(2)}$ which has
appeared before in \cite{desermc}. Comparing the terms linear in the sources in (\ref{lsd1}) and (\ref{sd2c}) we
arrive at the dual map between $\cL {SD}^{(1)}$ and $\cL {SD}^{(2)}$:

\be f^{\mu\nu}\leftrightarrow F^{\mu\nu}(f) \label{map1} \ee

\no Indeed, minimizing (\ref{sd2b}) at vanishing sources we find:

\be E_{\mu\alpha} \left( T_{\nu}^{\,\alpha}(f) - \eta_{\nu}^{\,\alpha} \frac {T(f)}2 \right) = - m \,
T_{\mu\nu}(f) \quad , \label{eq2} \ee

\no which can be recast as:

\be E_{\mu}^{\,\lambda} F_{\lambda\alpha} = m \left( \eta_{\mu\alpha} F - F_{\alpha\mu}\right) \quad .
\label{eq3} \ee

\no Comparing with (\ref{eq1}) we confirm the dual map (\ref{map1}) at classical level. The same map
\footnote{There is a mistake in the definition of the dual $F_{\mu\nu}$ on the right handed side of  formula
(19) of \cite{prd2009} where $T_{\mu\nu}$ should be replaced by $T_{\nu\mu}$.}  holds at quantum level up to
contact terms in the correlation functions. Since contact terms have no poles, the particle content of $\cL
{SD}^{(1)}$ and $\cL {SD}^{(2)}$ coincide, namely, one massive mode of helicity $+2$. From the master action
point of view this is a consequence of using a first-order Chern-Simons term ($CS_1$), which has no particle
content, as a mixing term in going from $\cL {SD}^{(1)}$ to $\cL {SD}^{(2)}$ \cite{prd2009}.

\subsection{Embedding $S_{SD}^{(2)}$ in $S_{SD}^{(3)}$}

It turns out that the Einstein-Hilbert term in (\ref{sd2c}) depends only upon the symmetric combination
$f_{(\mu\nu)}$, therefore it is  invariant under the local symmetry:

\be \delta_{\Gamma}
f_{\mu\nu}=\epsilon_{\mu\nu\alpha}\Gamma^{\alpha} \quad
,\label{sim2}\ee

\no which is broken by the first-order Chern-Simons mass term in (\ref{sd2c}). This suggests another round of
the NGE procedure. The Euler tensor from $S_{SD_2}^{(2)}$ is given by

\be M^{\mu\nu}=\frac{\delta S_{2}^{(2)}}{\delta
f_{\mu\nu}}=E^{\beta\mu}T^{\nu}_{\s\beta}-\frac{1}{2}E^{\nu\mu}T-mT^{\mu\nu}+G^{\mu\nu}(j).\ee

\no where we have defined

\be
G^{\mu\nu}(j)=\frac{E^{\lambda\mu}j^{\nu}_{\s\lambda}}{m}-\frac{E^{\nu\mu}j}{2m}.\ee

\no Again, with the help of an auxiliary field $a_{\beta\gamma}$ such that

\be \delta_{\Gamma} a_{\mu\nu} = - \epsilon_{\mu\nu\delta}\Gamma^{\delta} = -\delta_{\Gamma} f_{\mu\nu}
\label{dag}\ee

\no we can write the first iterated action:

\be S_1 = \int d^3\, x \left\lbrack \cL
{SD}^{(2)}+f_{\mu\nu}G^{\mu\nu}(j)+a_{\mu\nu}M^{\mu\nu}+{\cal{O}}(j^2)
\right\rbrack \label{s1b} \ee

\no such that

 \be \delta_{\Gamma} S_1 = \delta_{\Gamma} \left(- \int d^3\, x
\frac{m}{2}(a_{\mu\nu}E^{\mu\beta}a_{\beta}^{\s\nu})\right) \label{ds1b} \ee

\no So we derive

\be \cL 2= \cL {SD}^{(2)}+a_{\mu\nu}M^{\mu\nu}
+\frac{m}{2}(a_{\mu\nu}E^{\mu\beta}a_{\beta}^{\s\nu})
+f_{\mu\nu}G^{\mu\nu}(j)+{\cal{O}}(j^2),\label{l2b}\ee

\no which is invariant under (\ref{sim2}) altogether with (\ref{dag}). Although the equations of motion of the
auxiliary fields are not algebraic as in the last subsection, they can still be eliminated in a trivial way
leaving us with a local gauge invariant action. For this aim, note that the Euler tensor can be written as

\be
M^{\mu\nu}=E^{\mu\beta}\left(-T^{\nu}_{\s\beta}+\frac{\eta^{\nu}_{\beta}}{2}T+
m f_{\beta}^{\s\nu}-\frac{j^{\nu}_{\s\beta}}{m}
+\frac{\eta^{\nu}_{\beta}j}{2m}\right)\equiv
E^{\mu\beta}b_{\beta}^{\s\nu}\label{b}\ee

\no Now we can decouple the auxiliary fields by using:

\be a_{\mu\nu}E^{\mu\beta}b_{\beta}^{\,\nu}  +\frac{m}{2}(a_{\mu\nu}E^{\mu\beta}a_{\beta}^{\s\nu}) = -\frac
1{2m} b_{\mu\nu}E^{\mu\beta}b_{\beta}^{\,\nu} +
\frac{m}{2}(\tilde{a}_{\mu\nu}E^{\mu\beta}\tilde{a}_{\beta}^{\s\nu})\quad , \label{atil}\ee

\no Where $\tilde{a}_{\mu\nu}= a_{\mu\nu} + b_{\mu\nu}/m $. Neglecting the last term in (\ref{atil}) which has
no particle content, we obtain the invariant Lagrangian density:

\be \cL {{\rm inv.}} = \cL {SD}^{(2)} - \frac{1}{2m}b_{\mu\nu}M^{\mu\nu} +f_{\mu\nu}G^{\mu\nu}(j)\label{sd3a}\ee

\no Once again we have neglected quadratic terms in the sources. Although (\ref{sd3a}) is linear in the Euler
tensor, due to (\ref{b}) we have the general variation:

\be \delta S_{{\rm inv.}} = \int d^3 x \, M^{\mu\nu}\left( \delta f_{\mu\nu}-\frac{1}{m}\delta b_{\mu\nu}\right)
\label{ds2b}\ee

\no Consequently, the equations of motion of $S_{SD}^{(2)}$, $M^{\mu\nu}=0$, are embedded in the equations of
motion of $S_{{\rm inv.}}$. The action $S_{{\rm inv.}}$ is of third-order and can be rewritten, dropping terms
${\cal{O}}(j^2)$, as

\bea \cL {{\rm inv.}} \equiv \cL {SD}^{(3)} &=& -\frac{1}{2m}f_{\alpha\mu}(\Box
\theta^{\alpha\gamma}E^{\beta\mu}-\Box\theta^{\alpha\mu}E^{\beta\gamma}) f_{\gamma\beta} -
\frac{1}{2}T_{\mu\nu}T^{\nu\mu}+\frac{1}{4}T^{2}+j_{\mu\nu}\tilde{F}^{\mu\nu}(f)
\label{sd3b}\\
&=& -\frac{1}{8m}\left\lbrace \epsilon^{\mu\nu\rho}\Gamma_{\mu\gamma}^{\epsilon}\left\lbrack
\p_{\nu}\Gamma_{\epsilon\rho}^{\gamma} + (2/3)\Gamma_{\nu\delta}^{\gamma}\Gamma_{\rho\epsilon}^{\delta}
\right\rbrack\right\rbrace_{ff} - \frac 12 \left(\sqrt{-g} R\right)_{ff} +j_{\mu\nu}\tilde{F}^{\mu\nu}(f)
\label{sd3c}\eea

\no Note the change of the sign of the Einstein-Hilbert term. This is similar to the change of the sign of the
Maxwell term in going from (\ref{proca}) to (\ref{podolsky}). The theory $\cL {SD}^{(3)}$ corresponds to the
quadratic truncation of the topologically massive gravity of \cite{djt}. Above, we have defined

\be \tilde{F}^{\alpha\beta}(f) = \frac{E^{\alpha\gamma}E^{\beta\lambda} f_{(\gamma\lambda)}}{m^2} \label{ftil}
\ee

\no Comparing (\ref{lsd1}) and (\ref{sd3c}) we find the dual map

\be f_{\alpha\beta}\leftrightarrow \tilde{F}_{\alpha\beta} \quad .
\label{map2} \ee

\no Once again, the dual map (\ref{map2}) holds at classical and
quantum level (up to contact terms), see \cite{prd2009}, which
assures the spectrum equivalence between $S_{SD}^{(3)}$ and
$S_{SD}^{(2)}$.

\subsection{New self-dual model for spin-2 particles}

Only the symmetric combination $f_{(\mu\nu)}$ appears in $S_{SD}^{(3)}$  , explicitly,

\be S_{SD}^{(3)}\lbrack j \rbrack = \int d^3x\,\,\left\lbrack -
\frac{1}{2m}f_{(\lambda\mu)}\Box\theta^{\lambda\alpha}E^{\mu\delta}f_{(\alpha\delta)}-\frac{1}{2}f_{(\lambda\mu)}E^{\lambda\delta}E^{\mu\alpha}f_{(\alpha\delta)}+
j_{\lambda\mu}\tilde{F}^{\lambda\mu}(f) \right\rbrack\label{TMGS}\ee

\no Once more the highest derivative term of the action contains an extra local symmetry not shared by the
remaining terms. Namely, the first term of $S_{SD}^{(3)}$ is invariant under the linearized Weyl
transformation\footnote{The third order gravitational Chern-Simons ($CS_3$) term is invariant under Weyl
transformations $\delta_w g_{\mu\nu} = 2 \phi \, g_{\mu\nu}$ which reduce to $\delta_w f_{\mu\nu}=\phi \,
\eta_{\mu\nu}$ when we truncate $CS_3$ to quadratic terms about a flat background}:

\be \delta_w f_{\mu\nu}=\phi \, \eta_{\mu\nu}\label{sim3} \ee

\no while this is not true for the Einstein-Hilbert term. By imposing this new symmetry we will arrive at yet
another self-dual model for spin-2 particles in $D=2+1$. We start with the Euler tensor

\be M^{\beta\gamma}= \frac{\delta S_{SD}^{(3)}\lbrack j \rbrack }{\delta
f_{\beta\gamma}}=E^{\beta\mu}E^{\gamma\nu}b_{(\mu\nu)} = M^{(\beta\gamma)} \label{euler3}\ee

\no where $b_{(\mu\nu)}$ is given by: \be b_{(\mu\nu)}= - \left\lbrack
f_{(\mu\nu)}+\frac{(\eta_{\nu}^{\delta}E_{\mu}^{\s\alpha}+\eta_{\mu}^{\delta}E_{\nu}^{\s\alpha})f_{(\alpha\delta)}}{2m}+\frac{j_{(\mu\nu)}}{m^2}\right\rbrack\label{bmn}\ee

\no Following the same steps of last examples we end up with the action

 \be S_2 = \int d^3 x \, \left\lbrack \cL {SD}^{(3)} + a_{(\beta\gamma)}
 E^{\beta\mu}E^{\gamma\nu}b_{(\mu\nu)}-\frac{1}{2}a_{(\beta\gamma)}E^{\beta\mu}E^{\gamma\nu}a_{(\mu\nu)}
\right\rbrack \quad , \label{s2d}\ee

\no  After decoupling  the auxiliary fields and neglecting a term of the Einstein-Hilbert form
$(-1/2)\tilde{a}_{(\beta\gamma)}E^{\beta\mu}E^{\gamma\nu}\tilde{a}_{(\mu\nu)}$ where $\tilde{a}_{(\beta\gamma)}=
a_{(\beta\gamma)} - b_{(\beta\gamma)}$, which has no propagating degree of freedom, we obtain :

\bea \cL {SD}^{(4)} &=& \cL {SD}^{(3)}
+\frac{1}{2}b_{(\beta\gamma)}E^{\beta\delta}E^{\gamma\alpha}b_{(\alpha\delta)}\label{sd4a}\\
&=&
\frac{1}{4m^2}f_{(\rho\sigma)}(2\Box^2\theta^{\rho\nu}\theta^{\sigma\mu}-\Box^2\theta^{\rho\sigma}\theta^{\mu\nu})f_{(\mu\nu)}
+\frac{1}{2m}f_{(\lambda\mu)}\Box\theta^{\lambda\alpha}E^{\mu\delta}f_{(\alpha\delta)}
\label{sd4b}\\
&-&\frac{j_{(\alpha\delta)}E^{\rho\alpha}\Box\theta^{\delta\sigma}f_{(\sigma\rho)}}{m^3} \label{sd4c} \eea

\no This is a new self-dual model for particles of helicity $+2$ (or $-2$ depending on the sign of the
third-order term). It corresponds to the sum of a third order gravitational Chern-Simons term $\cL {CS3} \equiv
\epsilon^{\mu\nu\rho}\Gamma_{\mu\gamma}^{\epsilon}\left\lbrack \p_{\nu}\Gamma_{\epsilon\rho}^{\gamma} +
(2/3)\Gamma_{\nu\delta}^{\gamma}\Gamma_{\rho\epsilon}^{\delta} \right\rbrack $ and the fine tuned curvature
square term of \cite{bht} at linearized level with appropriate coefficients, i.e.,

\be\cL {SD}^{(4)} = \frac{1}{2m^2}\left(R^{\mu\nu}R_{\mu\nu}-\frac{3}{8}R^2\right)_{ff}
+\frac{1}{8m}\left\lbrace\epsilon^{\mu\nu\rho}\Gamma_{\mu\gamma}^{\epsilon}\left\lbrack
\p_{\nu}\Gamma_{\epsilon\rho}^{\gamma} + (2/3)\Gamma_{\nu\delta}^{\gamma}\Gamma_{\rho\epsilon}^{\delta}
\right\rbrack\right\rbrace_{ff} + j_{(\alpha\beta)}G^{\alpha\beta}(f)\label{sd4}\ee

\no where

\be G_{\alpha\beta}(f) = - \frac {\Box}{2 m^3}\left\lbrack E^{\rho}_{\s\alpha }\theta^{\delta}_{\s\beta} +
E^{\rho}_{\s\beta }\theta^{\delta}_{\s\alpha} \right\rbrack f_{(\rho\delta)} \label{g} \ee

The new model is invariant under all local symmetries (\ref{sim1}),(\ref{sim2}) and (\ref{sim3}). Since both
quartic- and third-order terms of (\ref{sd4}) are invariant by the same set of gauge symmetries, the NGE
procedure naturally terminates. Comparing (\ref{lsd1}) and (\ref{sd4}) we have the duality between
$S_{SD}^{(1)}$ and $S_{SD}^{(4)}$ established by the dual map:

\be f_{\mu\nu} \leftrightarrow G_{\mu\nu}(f) \label{map3} \ee

\no By using the identities $E_{\nu\mu}\theta^{\mu\delta}= E_{\nu}^{\s\delta} \, , \,
E_{\nu\mu}E^{\mu\delta}=\Box\theta_{\nu}^{\s\delta}$ and $E^{\nu\alpha} E^{\lambda\delta} =
\left(\theta^{\nu\delta}\theta^{\alpha\lambda}- \theta^{\nu\lambda}\theta^{\alpha\delta}\right)$ it is easy to
derive from (\ref{g}) that $E^{\lambda}_{\s\gamma}G^{\mu\gamma}= -(\Box^2/2 m^3)\left(\theta^{\lambda\alpha}
\theta^{\mu\beta} + \theta^{\mu\alpha} \theta^{\lambda\beta} - \theta^{\lambda\mu} \theta^{\alpha\beta} \right)
f_{(\alpha\beta)}$. Consequently, the equations of motion of $S_{SD}^{(4)}$:

\be \Box^2 \left(\theta^{\lambda\alpha} \theta^{\mu\beta} + \theta^{\mu\alpha} \theta^{\lambda\beta} -
\theta^{\lambda\mu} \theta^{\alpha\beta} \right) f_{(\alpha\beta)} = \Box \left( \theta^{\lambda\alpha}
E^{\mu\beta} + \theta^{\mu\alpha} E^{\lambda\beta} \right) f_{(\alpha\beta)} \ee

\no can be recast as

\be E^{\nu}_{\s\mu} G^{\mu\lambda} = G^{\nu\lambda} \label{eq4} \ee

\no which is exactly of the form (\ref{eq1}) if we note, see (\ref{g}), the identity $G^{\mu}_{\,\mu} = 0$.
Consequently, the dual map (\ref{map3}) between $S_{SD}^{(1)}$ and $S_{SD}^{(4)}$ is verified at classical
level. In particular, if we apply the operator $E^{\alpha}_{\, \nu}$ on (\ref{eq4}) and use(\ref{eq4}) again we
obtain the Klein-Gordon equation $\left(\Box - m^2\right)G_{\alpha\beta} = 0 $. It is remarkable that {\it all}
necessary constraints to describe a spin-2 massive particle, i.e., $G^{\mu}_{\,\mu}=0=G_{\lbrack\mu\nu\rbrack}
\, , \, \p^{\mu}G_{\mu\nu}=0=\p^{\nu}G_{\mu\nu}$ follow now from trivial identities instead of dynamic equations
differently from the other three self-dual models. In this sense the $S_{SD}^{(4)}$ model is the most natural
description of spin-2 parity singlets in $D=2+1$.

Regarding the particle content of the $\sdf $ model at quantum
level, we turn again to a master action:

\be S_M \lbrack f , \tilde{f} \rbrack = \int d^3 x \, \frac 12 \left\lbrack \frac 1m \left(\cL {CS3}
\right)_{ff} - \left(\sqrt{-g}R\right)_{ff} - \frac 1m \left(\cL {CS3}\right)_{f-\tilde{f},f-\tilde{f}}
\right\rbrack \label{sm3a}\ee

 \no Using the notation of \cite{prd2009} we can
rewrite (\ref{sm3a}) as follows\footnote{In \cite{prd2009} we have defined $\int h \cdot g \equiv \int d^3 x \,
h^{\mu\nu}\epsilon_{\mu}^{\s\alpha\beta}\p_{\alpha}g_{\beta\nu}$ and used $\Omega^{\alpha}_{\s\lambda} (f) = -
\epsilon^{\alpha\beta\gamma}\lbrack \p_{\lambda}f_{\gamma\beta} + 2\, \p_{\gamma} f_{(\lambda\beta)}\rbrack $ as
defined in \cite{desermc} up to an overall sign.}:

\bea S_M \lbrack f , \tilde{f} \rbrack &=&  \frac 14 \int \left\lbrack -\frac{\Omega (f) \cdot d \Omega (f)}{m}
+ f \cdot d \Omega (f) + \frac{\Omega (f-\tf) \cdot d \Omega
(f-\tf)}{m}\right\rbrack \label{sm3b} \\
&=& \frac 14 \int \left\lbrack \frac{\Omega (\tf) \cdot d \Omega (\tf)}{m} - \frac{2\, \Omega (\tf) \cdot d
\Omega (f)}{m}+
f \cdot d \Omega (f) \right\rbrack \label{sm3c}\\
&=&  \frac 14 \int \left\lbrack -\frac{\Omega (\tf) \cdot d \Omega (\Omega(\tf))}{m^2} + \frac{\Omega (\tf)
\cdot d \Omega (\tf)}{m} + \left( f - \frac{\Omega (\tf)}{m}\right)\cdot d \Omega\left(f - \frac{\Omega
(\tf)}{m}\right)\right\rbrack \nn \\ \label{sm3d} \eea

\no After the shift $f\to f + \Omega (\tf)/m$ the first two terms of (\ref{sm3d}) correspond exactly to the new
$S_{SD}^{(4)}$ self-dual model as function of $\tf_{\mu\nu}$ while the last term is a pure Einstein-Hilbert
action depending only on $f_{\mu\nu}$. So the particle content of the master action (\ref{sm3a}) is the same of
$S_{SD}^{(4)}$. On the other hand, if we start from (\ref{sm3b}) and shift $\tf_{\mu\nu} \to \tf_{\mu\nu} +
f_{\mu\nu}$ we obtain the $S_{SD}^{(3)}\lbrack f \rbrack $ model and a pure (decoupled) linearized gravitational
Chern-Simons term with no particle content \cite{djtprl}. Therefore, it is clear that $S_{SD}^{(4)}$ and
$S_{SD}^{(3)}$ share the same spectrum, i.e., one massive physical particle of helicity $+2$. This can be
confirmed by a calculation of the sign of the imaginary part of the residues at the poles of the  propagator of
$S_{SD}^{(4)}$ when saturated with conserved and traceless (as required by the linearized Weyl symmetry)
sources. In fact, there are two poles in the propagator of the $S_{SD}^{(4)}$ model, one massive and one
massless (ghost-like). It turns out that the traceless condition on the sources gets rid of a ghost-like
massless pole  and we are left with one physical massive pole \cite{dd}.

\section{Conclusion and comments}

In the second section we have shown that the (linearized) BHT model can be obtained from the Fierz-Pauli theory
via Noether embedment. In principle, the same procedure applies in higher dimensions however, the
Einstein-Hilbert action becomes dynamical for $D>3$ and by the arguments given here we expect that the embedment
would lead to a massless ghost. In other words, for $D>3$  the NGE of spin-2 massive particles is similar to the
spin-1 case where we have obtained a (``wrong'' sign) Maxwell-Podolsky theory.

The section 3 contains a natural chain of Noether gauge embedment: $S_{SD}^{(1)} \to S_{SD}^{(2)} \to
S_{SD}^{(3)} \to S_{SD}^{(4)}$. All terms of the  $S_{SD}^{(4)}$ model have the same local symmetry, so the
embedment terminates at the $S_{SD}^{(4)}$. This is similar to the spin-1 case where both Maxwell and firs-order
Chern-Simons terms are invariant under the same gauge transformations, so the embedment of the first-order model
of \cite{tpn} terminates after one round at the MCS theory.

It is interesting to remark that only in the model $S_{SD}^{(4)}$
the necessary constraints to describe a spin-2 particle are
identically (non-dynamically) satisfied which make us believe that
$S_{SD}^{(4)}$ is the most natural description of spin-2 parity
singlets in $D=2+1$, just like the MCS theory automatically
incorporates the transverse condition $\p_{\mu}F^{\mu}=0$, where
$F_{\mu}=\epsilon_{\mu\nu\alpha}\p^{\nu}A^{\alpha}/m $ is the dual
of the self-dual field $f_{\mu}$ of \cite{tpn}. Quite
surprisingly, the $S_{SD}^{(4)}$ model, which contains only third-
and fourth-order terms, is spectrally equivalent to the other
lower-order self-dual models. From the master action point of view
this follows from the triviality (no particle content) of the
linearized third-order gravitational Chern-Simons term ($CS_3$).
In fact, from this standpoint, the existence of the dual theories
$S_{SD}^{(2)}$, $S_{SD}^{(3)}$ and $S_{SD}^{(4)}$ follows from the
trivial cohomology of the differential operators which appear in
the $CS_1$, linearized Einstein-Hilbert and linearized $CS_3$
terms. There seems to be a one-to-one correspondence between
differential operators of trivial cohomology and dual theories.
This is also true in the spin-1 case where the  first-order
topological Chern-Simons term ($CS_1$) is apparently the only one
which could be used in the master action approach to generate a
dual theory to the first-order self-dual model of \cite{tpn}, in
this case one obtains the  Maxwell-Chern-Simons theory of
\cite{djt}.

Finally, since  $S_{SD}^{(4)}$ may be interpreted as the linearized version of the model $\cL {HDTMG}^{\pm} =
\pm \epsilon^{\mu\nu\rho}\Gamma_{\mu\gamma}^{\epsilon}\left\lbrack \p_{\nu}\Gamma_{\epsilon\rho}^{\gamma} +
(2/3)\Gamma_{\nu\delta}^{\gamma}\Gamma_{\rho\epsilon}^{\delta} \right\rbrack /8m + \sqrt{-g}\left\lbrack
R_{\mu\nu} R^{\nu\mu} - 3\, R^2/8 \right\rbrack /2m^2 $, one might consider it as a toy model for a massive
gravitational theory despite the absence of the Einstein-Hilbert term.

\vfill\eject

\centerline{{\bf Note added}}

After uploading our work we have been informed of the preprint \cite{andringa} where the $S_{SD}^{(4)}$ model
also appears.

\section{Acknowledgements}

D.D. is partially supported by \textbf{CNPq} while E.L.M. is supported by \textbf{CAPES}. We thank Dr. O. Hohm
for bringing \cite{andringa} to our knowledge.


\begin{thebibliography}{99}


\bibitem{djt} S. Deser, R. Jackiw and S. Templeton,
Annals Phys.140:372-411,1982, Erratum-ibid.185:406,1988, Annals Phys.281:409-449,2000.

\bibitem{tpn} P.K. Townsend, K. Pilch and P. van Nieuwenhuizen, Phys. Lett B {\bf 136} (1984)38.

\bibitem{dj} S.Deser and R. Jackiw, Phys.Lett.B {\bf 139} (1984)
371.

\bibitem{baeta} A.P. Baeta Scarpelli, M. Botta Cantcheff, J.A.
Helayel-Neto, Europhys.Lett.65:760-765,2004.

\bibitem{anacleto} M.A. Anacleto, A. Ilha , J.R.S. Nascimento, R.F. Ribeiro, C. Wotzasek,
Phys.Lett.B504:268-274,2001.

\bibitem{bht} E. Bergshoeff, O. Hohm and P.K. Townsend, Phys.Rev.Lett.102:201301,2009.

\bibitem{aragone} C. Aragone and A. Khoudeir, Phys. Lett.
B{\bf173} 141 (1986).

\bibitem{desermc} S. Deser and J. McCarthy, Phys. Lett. B{\bf 246}
441 (1990).

\bibitem{prd2009} D.Dalmazi and E.L.Mendon\c ca, Phys.Rev.D\textbf{79} 045025 (2009).

\bibitem{solda2} D.Dalmazi and E.L.Mendon\c ca, Phys.Rev.D\textbf{80} 025017 (2009).

\bibitem{podolsky} B.Podolsky, Phys.Rev.62:68-71,1942;  B.Podolsky and C. Kikuch, Phys.Rev.65:228-235,1944.

\bibitem{accioly} A.J. Accioly and H. Mukai, Z.Phys.C 75:187-191,1997.

\bibitem{jhep1} D. Dalmazi, JHEP 0601:132,2006, hep-th/0510153.

\bibitem{jhep2} D. Dalmazi, JHEP 0608:040,2006, hep-th/0608129.

\bibitem{fierz} M. Fierz, Helv. Phys. Acta {\bf 12} (1939) 3; M.
Fierz, W. Pauli, Proc. Roy. Soc. {\bf 173} (1939) 211.

\bibitem{oda} M. Nakazone and I. Oda, Prog. Theor. Phys. 121 (2009), 1389, see also arXiv:0902.3531.

\bibitem{gaitan}R. Gaitan, ``On the Coupling Problem of Higher Spin Fields in 2+1 Dimension '',
 PhD thesis, in spanish, arXiv:0711.2498.

\bibitem{djtprl} S. Deser, R. Jackiw and S. Templeton, Phys.Rev.Lett.48:975-978,1982.

\bibitem{dd} D. Dalmazi, ``Unitarity of spin-2 theories with linearized Weyl symmetry in $D=2+1$'',
arXiv:0908.1954.

\bibitem{andringa}  R. Andringa, Eric A. Bergshoeff, M. de Roo, O. Hohm, E. Sezgin and P. K. Townsend, `` Massive 3D Supergravity
'' arXiv:0907.4658.

\end{thebibliography}
\end{document}